\documentclass[%
 reprint,
 amsmath,amssymb,
 aps,
 showkeys,
]{revtex4-2}

\usepackage{graphicx}% Include figure files
\usepackage{epstopdf}
\usepackage{dcolumn}% Align table columns on decimal point
\usepackage{bm}% bold math
\begin{document}

\preprint{APS/123-QED}

\title{Strong constraints on Lorentz violation using new $\gamma$-ray observations around PeV}% Force line breaks with \\
%\thanks{A footnote to the article title}%

\author{Liang Chen$^{1,2,3}$}
\author{Zheng Xiong$^{1,2,3}$}
\email{xiongzheng@ihep.ac.cn}
\author{Cong Li$^{1,3}$}
\author{SongZhan Chen$^{1,3}$}%
\author{HuiHai He$^{1,2,3}$}%
\affiliation{%
$^1$Key Laboratory of Particle Astrophysics, Institute of High Energy Physics, Chinese Academy of Sciences, 100049 Beijing, China
}%
\affiliation{%
$^2$University of Chinese Academy of Sciences, 100049 Beijing, China
}%
\affiliation{%
$^3$TIANFU Cosmic Ray Research Center, Chengdu, Sichuan,  China
}%

\date{\today}% It is always \today, today,
             %  but any date may be explicitly specified

\begin{abstract}
The tiny modification of dispersion relation induced by Lorentz violation (LV) is an essential issue in quantum gravity (QG) theories, which can be magnified into significant effects when dealing with astrophysical observations at high energy and long propagation distance. LV would lead to photon decay at high energies, therefore, observations of high-energy photons could constrain LV or even QG theories. The Large High Altitude Air Shower Observatory (LHAASO) is the most sensitive gamma-array instrument currently operating above 100 TeV. Recently, LHAASO reported the detection of 12 sources above 100 TeV with maximum photon energy exceeding 1 PeV. According to these observations, the most stringent restriction is achieved in this paper, i.e., limiting the LV energy scale to $1.7\times10^{33} $ eV, which is over 139,000 times that of the Planck energy and achieve an improvement of about 1.9 orders of magnitude over previous limits.
\end{abstract}

%Use showkeys class option if keyword
                              %display desired
\maketitle

%\tableofcontents

\section{\label{sec:1}Introduction}

Three of the four fundamental interactions are described within quantum field theory framework as elementary particles in Standard Model (SM). Current understanding of the fourth interaction, gravity, is based on General Relativity (GR) which is well-confirmed in observation. However, GR cannot be reconciled with SM in extreme astrophysical objects, such as black hole \cite{penington2019replica}. A correct and consistent unification of SM and GR is one of the Holy Grail of modern physics, which has stimulated many theoretical ideas towards quantum gravity (QG) that describes gravity in the regimes where quantum effects cannot be disregarded. 

Several approaches to QG open fascinating perspectives on the structure of space-time, including string theory \cite{kostelecky1989spontaneous}, 
space-time foam \cite{Amelino_Camelia_1997,Amelino_Camelia_1998,li2021light}, 
non-commutative geometry \cite{Carroll_2001,amelino2000waves}, 
and loop QG \cite{Alfaro_2000,Alfaro_2003}. 
Some of them predict Lorentz violation (LV) at high energies, where Lorentz symmetry of space-time breaks down. Generally, these QG effects are assumed to be explicit while it approximates to Planck energy, $E_{QG}\sim E_{Pl}\equiv\sqrt{\hbar c^5/G}\simeq1.22\times10^{28}$ eV \cite{misner1973gravitation}. There is an enormous energy gap between $E_{Pl}$ and the highest energy particles known, the trans-GZK cosmic rays in $10^{20} $ eV \cite{Bahcall_2003}, which precludes any Earth-based experiment and direct observation of LV in $E_{Pl}$.

Fortunately, LV occurs in $E_{Pl}$ likely introduces some tiny ``LV relics" at relatively lower energies \cite{Liberati_2009}. And these ``relic" effects can demonstrate themselves and modify energy-momentum relation with additional terms suppressed by $E_{Pl}$. 

The modified dispersion relations (MDR) induced by LV is an essential issue in QG theories, mainly because it can be magnified into significant effects with high energy and long-baseline propagation. And these effects provide some available scenarios to validate these theories through ``windows on QG" \cite{Jacobson_2006,Liberati_2009}. Photons originated from remote energetic astrophysical objects are extremely promising to probe ``windows on QG" in various aspects:

\begin{itemize}
\item
  Anomalous threshold reactions induced by LV terms (photon decay, vacuum Cherenkov radiation);
\item
  Cumulative effects with long-baseline propagation (photon flight of time lags);
\item
  LV induced decays not characterized by a threshold (photon splitting).
\end{itemize} 

In the 1990s, Amelino-Camelia has suggested using Gamma-Ray Bursts (GRBs) to verify LV effects and LV limit, $E_{LV}$ \cite{Amelino_Camelia_1997,Amelino_Camelia_1998}. Based on observations of GRBs, Fermi-LAT restricted $E^{(1)}_{LV}$ to $9.3\times10^{28}$ eV in 2013  \cite{Vasileiou_2013}. And the High-Altitude Water Cherenkov (HAWC) Observatory finds evidence of 100 TeV photon emission from at least four astrophysical sources, which set a more stringent $E^{(1)}_{LV}$ to $2.2\times10^{31}$ eV last year \cite{Albert_2020}.

%The Large High Altitude Air Shower Observatory (LHAASO), located at 4410 m above sea level in the high mountains of the Sichuan Province, China, is a hybrid extensive air shower array designed to study cosmic rays and gamma-rays in the TeV-EeV energy range.  LHAASO consists of three arrays: WCDA, KM2A and WFCTA. WCDA is water Cherenkov detector with a total area of 0.078 km$^2$, which is 4 times that of HAWC and mainly for gamma-ray observations around TeV. At energy above 20 TeV, gamma-ray observation is mainly covered by the KM2A array. KM2A consists of the surface scintillation counters and the underground muon detector array. Thanks to the large area (1.3 km$^2$) and very high power of suppression of the cosmic ray background, the sensitivity of KM2A at energies above 100 TeV outreaches the performances of all current or planned ground-based gamma-ray detectors by at least one order. This means KM2A is a potent tool to search LV signatures in gamma-rays.

Guaranteed by excellent energy resolution and background rejection power, the Large High Altitude Air Shower Observatory (LHAASO) is the most sensitive detector above few tens of TeV which can help to search for LV signatures at ultra-high energy (UHE). Even with only five-month data and half configuration, the achieved sensitivity of KM2A is much better than all previous observations at energy above 100 TeV \cite{aharonian2021observation}. Basing on nearly one-year observations, LHAASO collaboration recently revealed significant number of PeVatrons in our Galaxy with reporting 12 gamma-ray sources over 100 TeV and the maximum observational energy of 1.42$\pm$0.13 PeV \cite{aharonian_2021}. %This means KM2A is a potent tool to search LV signatures in gamma-rays. %Additionally, the energy spectra of the three brightest objects are also reported and the maximum energy range has exceeded any previous observation. 

Inspired by the HAWC's previous work \cite{Albert_2020}, we use spectra from three sources and an UHE single photon event reported by LHAASO collaboration to explore anomalous behavior of photons induced by LV, which yields even more stringent limits to LV energy scale $E_{LV}$ than before.

Section \ref{sec:2} gives a brief introduction on LV phenomenological description which is enough to illustrate our methods to restrict $E_{LV}$, independent of the particular form of Lorentz violating theory. Section \ref{sec:3} shows $E_{LV}$ derived from the LHAASO's observations. The comparison between different methods are discussed in section \ref{sec:4}. Finally, we present our conclusion in section \ref{sec:5}.

\section{\label{sec:2}Lorentz violation}

A complete physical theory must include dynamics, and there are many attempts to provide a comprehensive framework to compute interactions. Both SM and relativity can be considered as effective field theory (EFT). Thus, it is intuitive for EFT frameworks to embed LV effects via LV operator \cite{SHAO_2010,mattingly2005modern}. 

%%%%%%%%%%%%%%%%%%%%%%%%%%%%%%%%%%%%%%%%%%%%%%%%%%%%%%%%%%%%%%%%%%%%%%%%%%%%%%%%%

Although the dynamics of LV are poorly understood till now, it does not bother us to focus on the kinematics of LV. In most of QG models, LV can be induced through MDR that may describe photon behavior from high-energy or distant astrophysical objects. For a phenomenological uniform description in natural units ($c=\hslash=1$), we use following MDR for photons:

\begin{equation}
E_\gamma^2=p_\gamma^2[1+\xi_n(\frac{p_\gamma}{E_{Pl}})^n]
\end{equation}

where $n = 1$ (linear modification) and $n = 2$ (quadratic modification) are relevant to current gamma-ray astronomy observations. Corresponding to n-th order LV  correction, $\xi_n$ is tiny dimensionless LV coefficient and suppressed by $E_{Pl}^{-n}$. Notably, $\xi_n>0$ represents superluminal photon propagation, while $\xi_n<0$ corresponds to subluminal propagation.

Subluminal photon propagation can be derived from radiative corrections caused by any charged particle with non-zero LV operators with mass dimension of four \cite{satunin2018one}. Thus, we consider only superluminal photon propagation to avoid lengthy discussion on radiative corrections. Before focusing on two interesting scenarios for superluminal propagation, a mathematical conversion $\alpha_n=\xi_n/E_{Pl}^n$ is conducted by inducing  $\alpha_n$ to interpret n-th order $E^{(n)}_{LV}$ more intuitively:

\begin{equation}
E^{(n)}_{LV}=\alpha_n^{-1/n}
\label{eq:one}
\end{equation}

\subsection{\label{PD}Photon Decay}

The basic QED vertex which single photon decays into electron and positron ($\gamma\rightarrow e^+e^-$) is  kinematically forbidden by canonical energy-momentum conservation. The matrix element and decay rate $\Gamma_{\gamma\rightarrow e^+e^-}$ can be calculated with MDR indeed. Once photon decay process is allowed, the decay rate goes like energy $E_\gamma$ above threshold of $\alpha_n$ with final momenta are non-parallel. Then any photon which propagates over macroscopic distances must below the energy threshold, or it performs as a hard cutoff in gamma-ray spectra without any high-energy photons observed on Earth \cite{Rubtsov_2012,Mart_nez_Huerta_2017}. Therefore, it will establish the threshold of $\alpha_n$ for any order n when $m_e$ and $E_\gamma$ stand for electron mass and gamma-ray energy, respectively.

\begin{equation}
\alpha_n\leq\frac{4m_e^2}{E_\gamma^n(E_\gamma^2-4m_e^2)}
\label{eq:two}
\end{equation}

Then, following Eq. (\ref{eq:one}) and (\ref{eq:two}), Mart\'inez-Huerta derived $a_0$, $E^{(1)}_{LV}$ and $E^{(2)}_{LV}$ with the LV generic approach \cite{Mart_nez_Huerta_2017}:

\begin{equation}
\alpha_0\leq\frac{4m_e^2}{E_\gamma^2-4m_e^2}
\end{equation}

\begin{equation}
E_{LV}^{(1)}\geq9.57\times10^{23}eV(\frac{E_\gamma}{TeV})^3
\end{equation}

\begin{equation}
E_{LV}^{(2)}\geq9.78\times10^{17}eV(\frac{E_\gamma}{TeV})^2
\end{equation}

\subsection{\label{PS}Photon Splitting}

The process that a photon splits into multiple photons ($\gamma \rightarrow N\gamma$) is allowed by energy-momentum conservation, but does not occur in standard QED because the matrix element and the phase space volume vanish. After introducing MDR, the matrix element and the phase space volume are non-zero, leading this non-threshold process occurs with a finite rate in superluminal photon propagation case. However, this process could provide limited contribution because splitting into more final-state photons is suppressed by more powers of the fine-structure  constant. Thus, the widest channel with quadratic modification is  three-photon channel ($\gamma\rightarrow 3\gamma$) \cite{Jacobson_2003,Maccione_2008}. Then we have photon splitting decay rate $\Gamma_{\gamma\rightarrow3\gamma}$ that is measured in unit of energy \cite{Astapov_2019,satunin2019new,Rubtsov_2017}.

\begin{equation}
\Gamma_{\gamma\rightarrow3\gamma}=5\times10^{-14}\frac{E_\gamma^{19}}{m_e^8 E_{LV}^{(2)\ 10}}
\end{equation}

Parallel with the concept of optical depth, the survival probability $P(E_\gamma,L_{obs};E_{LV})$ for photons after traveling distance $L_{obs}$ from astrophysical sources to Earth obeys an exponential distribution.

\begin{equation}
P(E_\gamma,L_{obs};E_{LV})=e^{-\Gamma_{\gamma\rightarrow3\gamma}\times L_{obs}}
\end{equation}

The free path of photon propagation $\lambda$ will fall with the increase of $E_\gamma$ given $E_{LV}^{(2)}$. The universe with the LV effect will become too opaque to observe gamma-rays traveling in distance of $L_{obs}$. Therefore, if we find the evidence of spectra energy distribution (SED) cutoff from energetic astrophysical sources in $L_{obs}$, we can infer that $\Gamma_{\gamma\rightarrow3\gamma}\times L_{obs} = 1$ where $L_{obs}$ is the critical value of $L_{LV}$. Then we can restrict the $E_{LV(3\gamma)}^{(2)}$:

\begin{equation}
E_{LV(3\gamma)}^{(2)}\geq3.33\times10^{19}eV(\frac{L_{obs}}{kpc})^{0.1}(\frac{E_\gamma}{TeV})^{1.9}
\label{eq:nine}
\end{equation}

%%%%%%%%%%%%%%%%%%%%%%%%%%%%%%%%%%%%%%%%%%%%%%%%%%%%%%%%%%%%%%%%%%%%%%%%%%%%%%%%%%%%%%%%%%%%%%%%%%%%%%%%%

\section{\label{sec:3}Constraints on Lorentz violation}

\subsection{\label{cutoff}Constraints based on spectra cutoff} 

In this analysis, we used the information of three luminous sources reported by LHAASO (LHAASO J2226+6057, LHAASO J1908+0621 and LHAASO J1825-1326) reported in \cite{aharonian_2021}. These spectra prefer log-parabolic fits. Even though there is a sign of steepening of spectrum with energy, there is no indication of sharp cutoff behavior. Additionally, the energy resolution of detector for gamma-ray events varies from different zenith angles. For showers with zenith angle less than $20^{\circ}$, the energy resolution is about 13\% at 100 TeV\cite{aharonian2021observation}. So we performed $\chi^2$-fitting to log-parabolic spectra for these sources convolved with a moderate estimation on energy resolution of 20\%. The $\Delta\chi^2$ between fitting with cutoff at $E_{cut}$ and that without cutoff is calculate:

\begin{equation}
\Delta\chi^{2} = \chi^{2}(E_{cut})-\chi^{2}(E_{cut}\rightarrow\infty)
\end{equation}

As the result of preference to non-cutoff behavior, the $\Delta\chi^2$ decreases with the increase of energy. So we proceed ti set one-side lower limit on $E_{cut}$, which 95\% CL corresponds to  $\Delta\chi^2 = 2.71$. The achieved lower limits on $E_{cut}$ for the three sources are listed in TABLE \ref{tab:tablelong}. The highest limit, 370.5 TeV, is from LHAASO J1908+0621. 

\begin{table*}[t]
\caption{\label{tab:tablelong}LHAASO sources with the most conservative distance of the possible origin. We calculate the 95\% CL lower limits for $E_c$ and $E_{LV}^{(n)}$, upper limits for $\alpha_0$, respectively.}
\begin{ruledtabular}
\begin{tabular}{ccccccc}
 Source&$E_{cut}$ (TeV)&$L_{obs}$(kpc)&$\alpha_0(10^{-18})$&$E^{(1)}_{LV}(10^{31}$eV)
&$E^{(2)}_{LV}(10^{22}$eV)& $E^{(2)}_{LV(3\gamma)}(10^{24}$eV)\\ \hline
LHAASO J2226+6057 & 280.7 & 0.8\footnote{The distance of possible astrophysical objects associated with LHAASO J2226+6057 is given in \cite{Kothes_2001}.}& 13.26 &2.11 & 7.71 &1.46\\
LHAASO J1908+0621 & 370.5 & 2.37& 7.61 & 4.87 & 13.43 & 2.76 \\
LHAASO J1825-1326 & 169.9 & 1.55 & 36.18 & 0.47 & 2.82 & 0.60 \\\hline
Combined sources & 483.3 & - & 4.47 & 10.80 & 22.84 & 4.37\\
\end{tabular}
\end{ruledtabular}
\end{table*}

\begin{figure}[t]
\includegraphics[height=60mm,width=80mm]{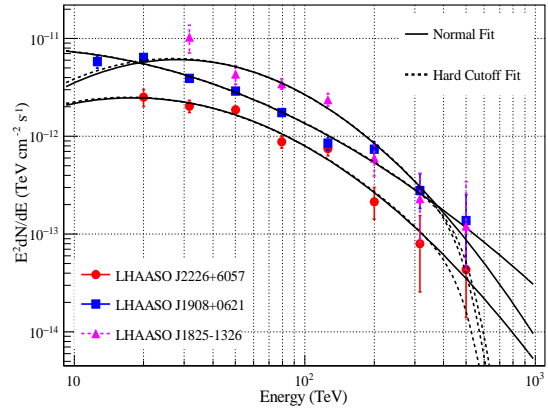}% Here is how to import EPS art
\caption{\label{fig:sed}Comparison of the normal best-fit spectra with those with a hard cutoff at   $E_{cut}$=483.3 TeV. The fitting lines have taken account of 20\% energy resolution of LHAASO. }
\end{figure}

\begin{figure}[t]
\includegraphics[height=60mm,width=80mm]{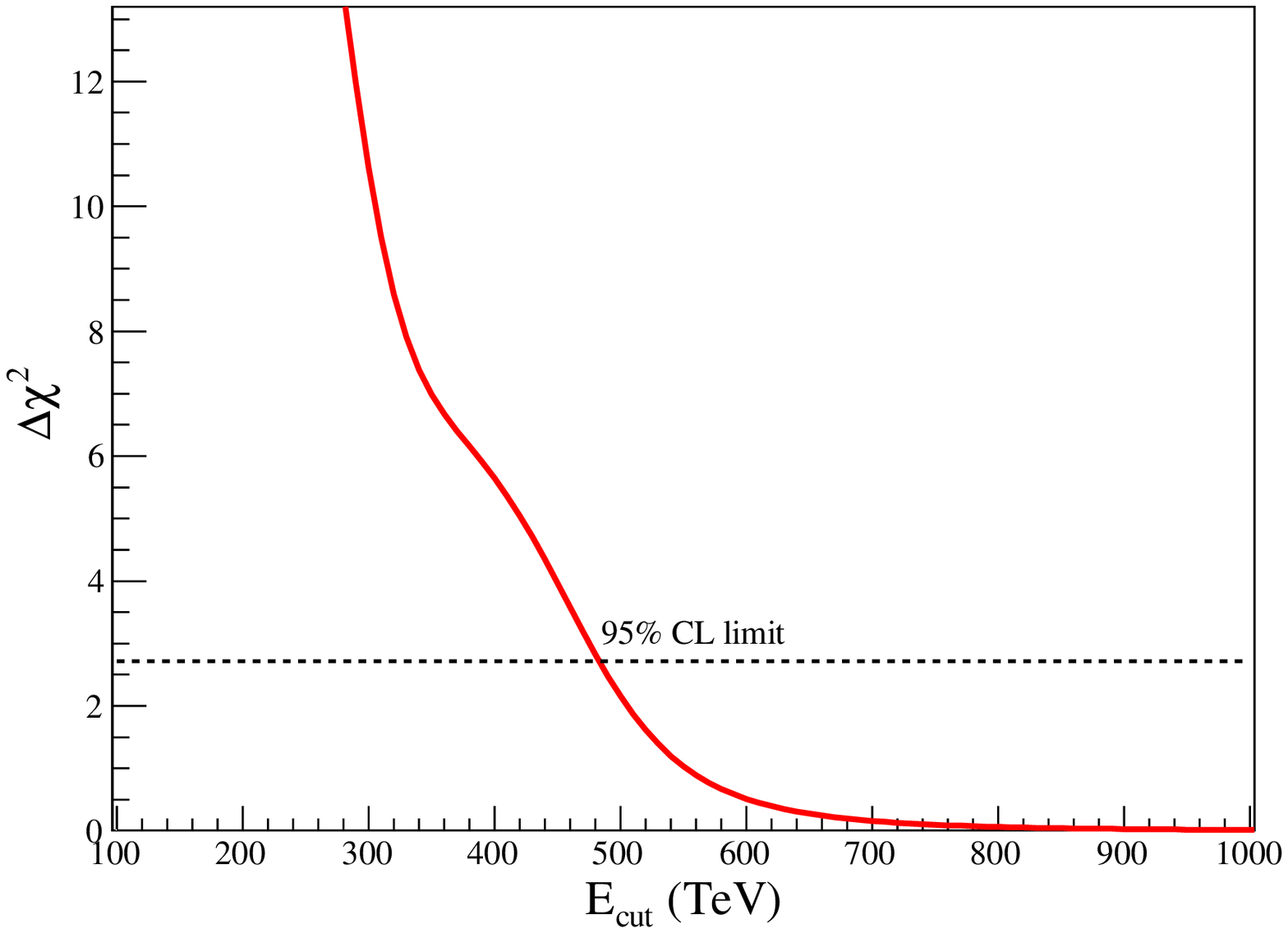}\\% Here is how to import EPS art
\includegraphics[height=60mm,width=80mm]{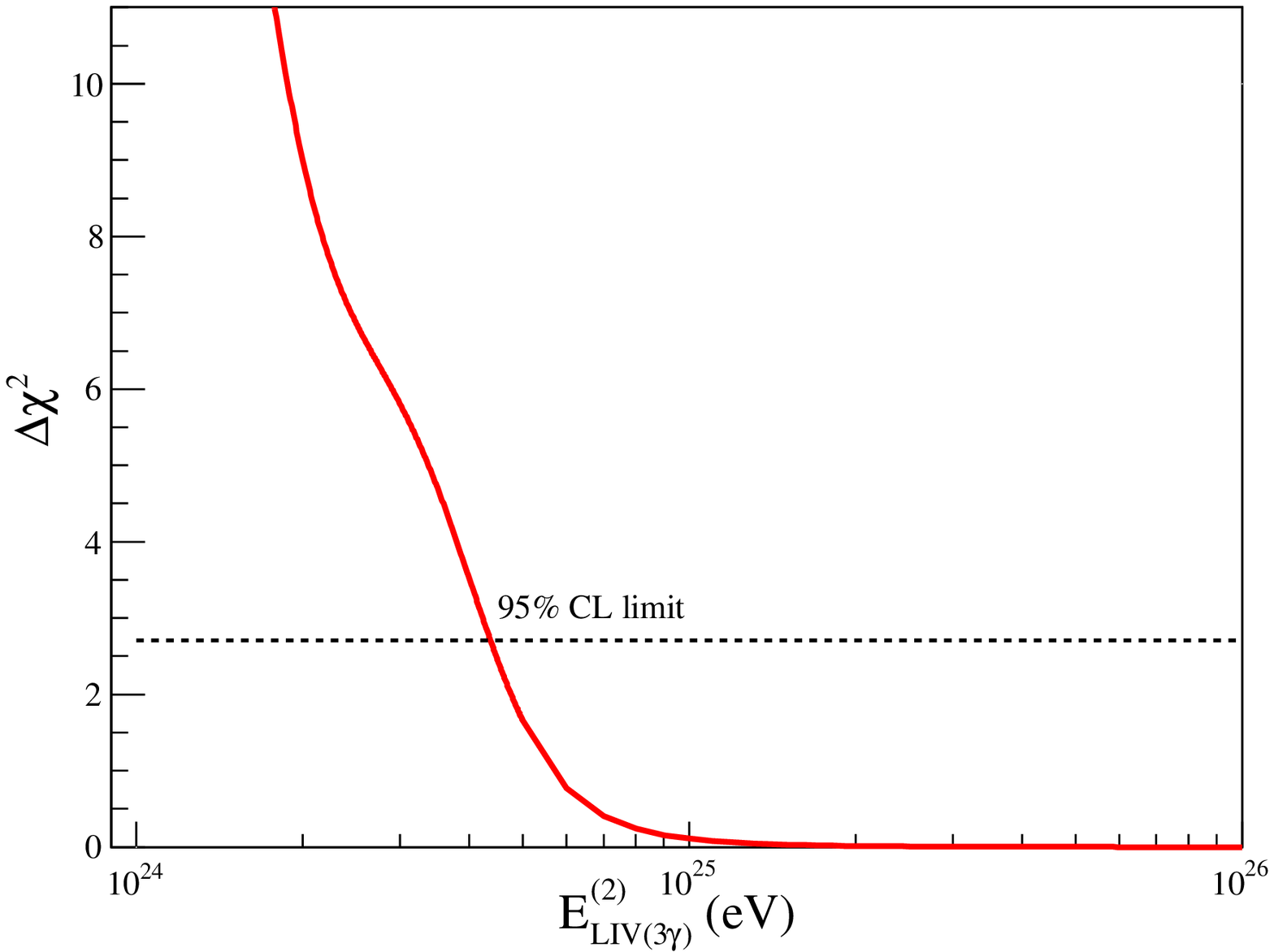}
\caption{\label{fig:chi} In combined analysis, the $\Delta\chi^2$ is a function of $E_{cut}$ or $E^{(2)}_{LIV(3\gamma)}$ and the dotted line indicate the 95\% CL Lower limit. Up: Photon decay scenario. Down: Photon splitting scenario.}
\end{figure}

In photon decay scenario, the spectra of any astrophysical sources would perform the cutoff behavior at the same energy if these sources are not limited by their acceleration mechanism, so we can accumulate the three sources into the combined statistic estimator to conduct the combined analysis. We also compared the best-fitting spectra before and after using the same cutoff shown in FIG.\ref{fig:sed}. The achieved $95\%$ CL lower limit for $E_{cut}$ is improved about 30\% to 483.3 TeV after combined analysis. In combined analysis, the $\Delta\chi^2$ is a function of $E_{cut}$ shown in FIG. \ref{fig:chi}. And the most stringent limits on $E_{LV}$ are well above $E_{Pl}$, which is up to $4.87\times10^{31}$ eV using single source and $1.08\times10^{32}$ eV using combined analysis in linear modification. In quadratic modification, $E^{(2)}_{LV}$s are still far below $E_{Pl}$. 

\begin{figure}[t]
\includegraphics[height=60mm,width=80mm]{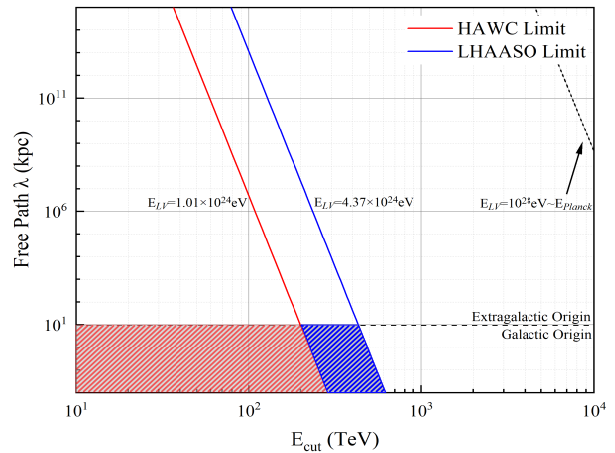}% Here is how to import EPS art
\caption{\label{fig:PS}In photon splitting scenario, the relation between free path $\lambda$ and spectra cutoff energy $E_{cut}$ are plotted with solid lines, given different $E_{LV}$s which are set by HAWC, LHAASO and $E_{Pl}$.}
\end{figure}
 
In photon splitting scenario, the expected $E_{cut}$ is distance dependent, but $E_{LV}$ would be still universal energy scale for different sources where Lorentz symmetry breaks down. To conduct the combined analysis, we used the same $E_{LV}$ rather than $E_{cut}$ to add the three sources into the combined statistic estimator. In combined analysis, the $\Delta\chi^2$ is a function of $E_{LV(3\gamma)}^{(2)}$ in FIG. \ref{fig:chi}. The deduced lower limits on $E_{LV}$ are listed in TABLE \ref{tab:tablelong}. We use the nearest distance among possible origins to set most conservative $E_{LV}$. Besides the most stringent limits on $E_{LV}$ are up to $2.76\times10^{24}$ eV using single source and $4.37\times10^{24}$ eV using combined analysis in quadratic modification. The results with quadratic modification are far blow $E_{Pl}$. Actually, the conservative distance seems underestimate the lower bound of $E_{LV}$. Only when we confirm lots of galactic Pevatrons, can we improved $E_{LV}$. 

However, it should be noted that photon splitting induced by LV is a non-threshold process as we mentioned in Sec \ref{sec:1} and Sec \ref{PS}. Thus, it is necessary to clarify that the photon splitting induced by LV derives ``quasi-threshold" for spectra cutoff behavior within galactic astrophysical sources on the kpc scale. To illustrate this issue, we used Eq.(\ref{eq:nine}) to calculate the relation between free path $\lambda$ of photon propagation and spectra cutoff energy $E_{cut}$ with given $E_{LV}$s in FIG.\ref{fig:PS}. The free path $\lambda$ shrinks exponentially with the increase of $E_{cut}$ in nineteenth power, so that photon splitting could be treated as a ``quasi-threshold" scenario. Considering TeV gamma-ray photons will interact with intergalactic diffuse photon background, it is a good choice to choose galactic astrophysical sources to examine the photon splitting scenario. The characteristic size scale of the Milky Way is marked in a black dash line. And the parameter space marked in the colored pattern is observable to set a limit to $E_{LV}$ within galactic distance in practice. Conversely, the higher $E_{LV}$ is, the more transparent universe will be for photons to travel. LHAASO achieved about 4 times $E_{LV}$ of the previous result report by HAWC based on spectra cutoff. And the energy gap between $E_{LV}$ set by LHAASO and $E_{Pl}$ remains to be tested in the future.

In addition, we quantified the effect on the fitting $E_{cut}$ with different energy resolutions, because we have no access to the simulation information of LHAASO. And $E_{cut}$ increases with the increment of energy resolution. The combined analysis is affected the most , but $E_{cut}$ in 95\% CL with 30\% energy resolution, $E_{cut}^{30\%}=467.7$ TeV. The uncertainty induced by energy resolution approximates to 3\%.

\subsection{\label{single}Constraint based on single event} 

While dealing with the phase space integration to calculate photon decay rate, the condition that the momentum of electron or positron should be real and positive which gives a stringent limit to $\alpha_n$ \cite{Mart_nez_Huerta_2017}. According to Eq. (\ref{eq:one}), the photon decay rate $\Gamma_{\gamma\rightarrow e^+e^-}$ will grow with $\alpha_n$ in fixed photon energy $E_{\gamma}$, final-state particles mass $m_e$ and the leading order of modification $n$, while $\alpha_n$ is above the threshold. If $\alpha_n$ is plenty big enough, $E_{LV}$ will approximate to $E_{\gamma}$. And photo decay process will become so efficient that the free path is limited to millimeter scale. In other words, there's tiny survival probability for photon during the propagation to earth when $E_{LV}$ is a few orders of magnitude higher than photon in this LV generic approach. If we find the evidence of UHE photons then $E_{LV}$  will be push higher than $E_{\gamma}$.

In spectral analysis, the bin width is always larger than energy resolution to ensure enough statistics, which is at the price of loss of energy information about a single event. In addition, the possibility of misidentifying a cosmic ray as a photo-like event is also not fully considered for spectral analysis. Thanks to excellent energy resolution and rejection power, LHAASO can distinguish the UHE photons among events.

Meanwhile, LHAASO does find evidence of a PeV single gamma-ray event from LHAASO J2032+4102 in the Cygnus region. The probability of non-rejection of a cosmic ray is estimated to be 0.028\% in \cite{aharonian_2021}. Thanks to LHAASO decent energy resolution above 100 TeV, this event's energy uncertainty is limited within $1.42\pm0.13$ PeV. Hence, we can be confident that this UHE photon's lower energy bound is 1.21 PeV in 95\% CL. This UHE single photon event can play as a counter example against low $E_{LV}$ assumptions. And $E^{(n)}_{LV}$ set by this event is given in TABLE \ref{tab:tableshort}.

\begin{table*}[t]%The best place to locate the table environment is directly after its first reference in text
\caption{\label{tab:tableshort}%
The UHE single photon event comes from LHAASO J2032+4102. The energy of this event reaches $1.42\pm0.13$ PeV and 95\% CL lower bound of $E^{95\%}_{\gamma,low}=1.21$ PeV. The $L_{obs}$ is $1.40\pm0.08$ and 95\% lower bound of  $L^{95\%}_{obs,low}=1.27$ kpc.}
\begin{ruledtabular}
\begin{tabular}{ccccccc}
\textrm{}&
\textrm{$E_{\gamma}$(PeV)}&
\textrm{$L_{obs}$(kpc)}&
\textrm{$\alpha_0(10^{-19})$}&
\textrm{$E^{(1)}_{LV}(10^{33}$eV)}&
\textrm{$E^{(2)}_{LV}(10^{24}$eV)}&
\textrm{$E^{(2)}_{LV(3\gamma)}(10^{25}$eV)}\\
\colrule
UHE event & 1.21  & 1.27\footnote{The distance of possible astrophysical objects associated with LHAASO J2032+4102 is given in \cite{Rygl_2012}. } & 7.13 & 1.70 & 1.43 & 2.45\\
\end{tabular}
\end{ruledtabular}
\end{table*}

%%%%%%%%%%%%%%%%%%%%%%%%%%%%%%%%%%%%%%%%%%%%%%%%%%%%%%%%%%%%%%%%%%%%%%%%%%%%%%%%%%%%%%%%%%%%%%%%%%%%%%%%%

\section{\label{sec:4}Discussion}

According to our calculation, the most stringent constraint on $E_{LV}$ through the UHE single photon event reported by LHAASO is up to $1.7\times10^{33}$ eV, which is over 139,000 times that of $E_{Pl}$. The limitation methods on $E_{LV}$ had a great development since Amelino-Camelia first suggested to use astrophysical photons to testify QG in 1998 \cite{Amelino_Camelia_1998}.

Regarding high-energy photons, it is a direct way to restrict $E_{LV}$ through measuring the arrive-time delay with long-baseline propagation but affected by uncertain emitting time and region from different luminous sources.

Several regression methods are used to search the flight of time delay under the assumption that all GRBs have the same emitting mechanism and similar emitting-time postpone. Due to this long-baseline propagation, we have to consider the absorption of extra-galactic background light (EBL) or cosmic microwave background (CMB), making it harder to detect GRBs' photons with energy above TeVs through satellite-borne observation. Hence, the LV constraints with the cumulative effects method may be confined to the energy range near $E_{Pl}$, i.e., $E_{LV}$ set by Fermi-LAT is 7.6 times over $E_{Pl}$\cite{Vasileiou_2013}.

Indeed, the methods based on non-cumulative effect can avoid this dilemma. It is more reasonable to set $E_{LV}$ by photon's collective behavior, for instance, cutoff behavior of SED. With the accumulation of high-energy events, SED will reach a higher energy range, leading to a more stringent constraint on $E_{LV}$. However, SED likely would show steepening or cutoffs at some energy because of acceleration and radiation mechanism. And the cutoff behavior of SED may be induced by photon propagation effects or absorption with EBL or CMB. Both of them will introduce uncertainty to $E_{LV}$ deduced by SED cutoff. So the LV constrains set by SED cutoff behavior is likely to be restricted within the energy range which is three or four orders of magnitude above $E_{Pl}$, i.e, HAWC and LHAASO set the $E^{(1)}_{LV}$ to $2.22\times10^{31}$ eV and $1.08\times10^{32}$ eV which are 1800 and 8850 times of $E_{Pl}$ respectively.

In this aspect, we can continue to set more stringent LV restrictions by an UHE single event. The only penalty we should pay is the uncertainty for failure to reject a cosmic-ray event. While this method is heavily grounded on background rejection power and energy resolution, LHAASO-KM2A is capable of rejecting cosmic ray background by a factor of $10^{-4}$ above 100 TeV and performs at the energy resolution of 13\% above 100 TeV according to \cite{aharonian_2021}. In short, LHAASO has competence in restricting the $E_{LV}$ through single-photon event in photon decay scenario and improve $E_{LV}$ to over five orders of magnitude over $E_{Pl}$.

%%%%%%%%%%%%%%%%%%%%%%%%%%%%%%%%%%%%%%%%%%%%%%%%%%%%%%%%%%%%%%%%%%%%%%%%%%%%%%%%%%%%%%%%%%%%%%%%%%%%%%%%%

\section{\label{sec:5}Conclusion}

In this work, we used the information of three sources above 100 TeV and an UHE single photon event from Cygnus region to set limits on $E_{LV}$. We have considered two scenarios of photon decay and photon splitting, among which the first-order photon decay process gives the most stringent Lorentz limit, $1.7\times10^{33}$ eV. The newly given limit is about 1.9 orders of magnitude over previous limits \cite{Albert_2020}. 

It is surprising that half-configured LHAASO-KM2A provides high quality data in nearly one-year operation. LHAASO-KM2A will stand out to restrict $E_{LV}$ by taking the advantages of the UHE photons from sources owning to its unprecedented capability of background rejection, high energy and angular resolutions. Hence LHAASO-KM2A is likely to perform as a unique probe to LV physics based on UHE single photon events in the future.

\begin{acknowledgments}
This work is supported in China by National Key R\&D program of China under the grant 2018YFA0404201, by National Science Foundation of China (NSFC) No.12022502.
%\dots.
\end{acknowledgments}

% The \nocite command causes all entries in a bibliography to be printed out
% whether or not they are actually referenced in the text. This is appropriate
% for the sample file to show the different styles of references, but authors
% most likely will not want to use it.
\nocite{*}

\bibliography{apssamp}% Produces the bibliography via BibTeX.

\end{document}